\documentstyle[prl,aps,multicol,epsfig]{revtex}

\begin{document}
\draft

\title{ Mott-Hubbard Transition in Infinite Dimensions}

\author{R.~M.~Noack}
\address{Institut de Physique Th\'eorique, Universit\'e de Fribourg,
CH-1700 Fribourg, Switzerland}
\author{F.~Gebhard}
\address{Fachbereich Physik, Philipps-Universit\"at Marburg, D-35032 
Marburg, Germany}
\date{{\bf Final version, Phys. Rev. Lett. 82, 1915 [1999]}}
\maketitle

\begin{abstract}
We calculate the zero-temperature gap and quasiparticle weight
of the half-filled Hubbard model with a random dispersion relation.
After extrapolation to the thermodynamic limit, we obtain reliable bounds
on these quantities for the Hubbard model in infinite dimensions.
Our data indicate that the Mott-Hubbard 
transition is continuous, i.e., that the quasiparticle weight becomes zero 
at the same critical interaction strength at which the gap 
opens.
\end{abstract}


\begin{multicols}{2}
\narrowtext

As pointed out by Mott~\cite{Mott} many years ago,
a sufficiently strong electron-electron interaction 
produces a gap for current-carrying charge excitations
in systems with an integer average number of electrons per lattice site.
In their simplest form, his ideas can be understood within
the one-band Hubbard model~\cite{Hubbard}, in which the kinetic energy
of the electrons competes with their purely local interaction.
When there is one electron per lattice site on average
and the interaction strength, $U>0$, is large compared to the bandwidth, $W$,
the charge gap is of the order of $\Delta(U\gg W)=U-W$, irrespective
of a possible symmetry breaking in the ground state.
Hence, in the absence of nesting, 
the Hubbard model at half band-filling contains a zero-temperature
quantum phase transition from a metal 
to the Mott-Hubbard insulator at some critical
interaction strength~$U_{\text{c}}$
of the order of the bandwidth, $U_{\text{c}}\approx W$~\cite{Mott}.

The Mott-Hubbard transition poses an intricate many-body
problem, and few exact results are available. 
In one dimension, the quantum phase transition can be investigated
in the Hubbard model with a linear dispersion relation in which the
absence of nesting precludes an antiferromagnetic instability at 
$U=0^+$.
In this model, the metal ceases to exist when the gap opens
at $U_{\text{c}}=W$~\cite{GRPRL},
i.e., the Mott transition is continuous~\cite{BUCH}.

The Hubbard model cannot be solved exactly in $d>1$ dimensions.
However, many simplifications arise in the limit 
$d\to\infty$~\cite{MVPRLdinfty}. 
In this limit,
the Hubbard model can be mapped onto a single-impurity problem
in a bath of electrons whose properties must be 
determined self-consistently (dynamical mean-field theory);
for recent reviews, see Refs.~\cite{PJF,MEGAREVMOD,BUCH}. 
From this mapping it can be shown that a metallic ground state must
be a Fermi liquid whose quasiparticle weight~$Z$ is identical
to the size of the jump discontinuity of the momentum 
distribution~\cite{MH}. 

\begin{figure}
\vspace*{-0.6cm}
  \begin{center}
    \epsfig{file=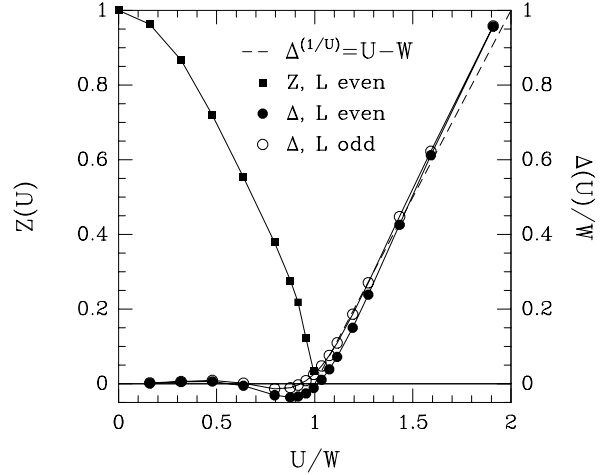,width=8cm}
  \end{center}
\vspace*{-0.2cm}
\caption{
  The quasiparticle weight $Z(U)$ and the
  one-particle gap, $\Delta(U)$, extrapolated to the thermodynamic
  limit separately for even and odd system sizes.
  }
\label{fig:Zgapinf}
\end{figure}

Within the dynamical mean-field theory, the interacting Green function must
be calculated explicitly at all frequencies in order to iterate the
dynamical mean-field equations.
In this work we avoid this self-consistency scheme altogether
by studying the Hubbard model with a random dispersion relation, 
which 
is equivalent to the Hubbard model in infinite dimensions~\cite{BUCH}.
We work directly with static ground-state properties such as the gap
and the quasiparticle weight (see Eqs.~(\ref{gapdef},~\ref{nkdef}) below) 
which can be
used to determine the nature
of the Mott-Hubbard metal-insulator transition.
We calculate these quantities using exact diagonalization on finite lattices
of up to $L=14$ sites, and extrapolate to the thermodynamic limit.
As shown in Fig.~\ref{fig:Zgapinf}, our extrapolated results for the
gap and the quasiparticle weight are
consistent with a continuous Mott--Hubbard transition at 
$U_{\text{c}}\approx W$ for a semielliptic density of states.

Our results are in strong disagreement with the scenario of a
preformed gap proposed by Georges, Kotliar {\it et al.}~\cite{MEGAREVMOD}.
They find a metallic solution of the self-consistency equations 
up to some $U_{\text{c},2}$, and an insulating solution down to
$U_{\text{c},1} < U_{\text{c},2}$,
with $U_{\text{c},2} - U_{\text{c},1} \approx 0.4 W$.
They argue that the transition occurs at 
$U_{\text{c}}= U_{\text{c},2}$ because the metallic solution has
lower energy.
The quasiparticle weight then vanishes
continuously, but the gap opens abruptly \cite{MEGAREVMOD}.
A coexistence regime at zero temperature implies
a line of first-order metal-insulator phase transitions 
at finite temperatures, and it is tempting to explain
the qualitative features of the phase diagram of
V$_2$O$_3$ solely on the basis of electron-electron
interactions~\cite{MEGAREVMOD}.
Recently, however,
the arguments presented in favor of 
a discontinuous transition have been
scrutinized and argued to be insufficient~\cite{David,BUCH,Kehrein}. 

The Hubbard model on an $L$-site lattice is given by
\begin{equation}
\hat{H} 
= \sum_{l,m=1;\sigma}^L t(l-m) \hat{c}_{l,\sigma}^+\hat{c}_{m,\sigma} +
U \sum_{l=1}^L \hat{n}_{l,\uparrow}\hat{n}_{l,\downarrow} \, ,
\end{equation}
where $\hat{c}_{l,\sigma}^+$ ($\hat{c}_{l,\sigma}$) creates (annihilates)
an electron with spin~$\sigma$ on lattice site~$l$,
$\hat{n}_{l,\sigma}=\hat{c}_{l,\sigma}^+\hat{c}_{l,\sigma}$
denotes the number operator on site~$l$, and
$t(r)=t^*(-r)$. 
The kinetic energy operator
is diagonal in momentum space,
$\hat{T}= \sum_{k,\sigma} \varepsilon_k 
\hat{c}_{k,\sigma}^+ \hat{c}_{k,\sigma}$,
$\varepsilon_k=(1/L) \sum_{r} t(r) \exp(-ikr)$.
Here, 
$\hat{c}_{k,\sigma}^+ =\sqrt{1/L} \sum_{l}\exp(ikl) \hat{c}_{l,\sigma}^+$ 
creates an electron with Bloch momentum~$k$.

Two important functions which characterize the dispersion relation
are the non-interacting density of states,
$D(\varepsilon) = (1/L) \sum_{k} \delta(\varepsilon-\varepsilon_k)$,
and the joint density of states,
$D_q(\varepsilon_1,\varepsilon_2) = (1/L) 
\sum_{k}\delta(\varepsilon_1-\varepsilon_k)
 \delta(\varepsilon_2-\varepsilon_{k+q})$.
The latter quantity gives the momentum-averaged probability for the event
that a momentum transfer~$q$ generates a transfer of kinetic energy 
from $\varepsilon_1$ to $\varepsilon_2$. 
In the random dispersion approximation (RDA)~\cite{BUCH}
the dispersion $\varepsilon_k$
is replaced by the random dispersion $\varepsilon^{\text{RDA}}_k$
with the same density of states, $D^{\text{RDA}}(\varepsilon)=D(\varepsilon)$,
but a completely uncorrelated joint density of states,
$D_{q\neq 0}^{\text{RDA}}(\varepsilon_1,\varepsilon_2) = 
D(\varepsilon_1)  D(\varepsilon_2)$.
All higher correlation functions must also 
factorize correspondingly.
In order to satisfy these conditions,
we randomly assign an energy $\varepsilon^{\text{RDA}}_k$ 
chosen from the probability distribution~$D(\varepsilon)$
to each momentum~$k$.
In this version of the RDA all forms of magnetic order 
are suppressed (``fully frustrated hopping''~\cite{MEGAREVMOD})
such that magnetic order does not conceal the Mott-Hubbard transition.

One way to see the equivalence of the RDA to the
limit of infinite lattice dimensions is the diagrammatic approach.
The value of the local {\sl non-interacting\/}
Green function in the RDA is of order unity, 
and identical to that of the Hubbard model in infinite dimensions
for a given~$D(\epsilon)$.
For all other lattice distances the Green function
behaves like $\sqrt{1/L}$~\cite{BUCH}.
Two vertices $l$ and $m$ of a skeleton diagram in the proper self-energy
are connected by {\it three\/} independent Green function 
lines~\cite{MVPRLdinfty} such that the contribution from $l-m \neq 0$ 
vanishes like $L{\cal O}(L^{-3/2})=
{\cal O}(L^{-1/2})$. Thus, the self-energy becomes purely local.
Moreover, the skeleton expansion
of the Hubbard model in infinite dimensions and in the RDA 
are identical. The local {\sl interacting\/} Green functions also agree 
to all orders in perturbation theory
since they depend on momentum only through~$\varepsilon_k$.
Thus, the self-energies of both approaches are the same,
i.e., they lead to the same mean-field theory.

For a numerical analysis we must specify
the dispersion relation $\varepsilon^{\cal Q}_k$ of a 
realization~${\cal Q}$ of the randomness,
which then defines the corresponding Hubbard Hamiltonian
$\hat{H}^{{\cal Q}} = \sum_{k,\sigma} \varepsilon^{\cal Q}_k
  \hat{c}_{k,\sigma}^+ \hat{c}_{k,\sigma} 
+ U \sum_{l} \hat{n}_{l,\uparrow}\hat{n}_{l,\downarrow}$.
Each $\hat{H}^{{\cal Q}}$ is 
a well-defined Hamiltonian on a formally
one-dimensional lattice which can be treated using numerical
techniques.
Here we use exact diagonalization with the Davidson 
algorithm~\cite{davidson} to obtain energies and equal-time matrix elements of
low-lying states which are accurate to close to machine precision.
Since we will have to average over configurations ${\cal Q}$ which
correspond to diagonal disorder in momentum space, we perform the
diagonalization in the momentum-space basis.
We adopt antiperiodic boundary conditions for an even number
of sites~$L$ and periodic boundary conditions for odd~$L$.

For a given~$U$,~$L$, we 
calculate the single-particle gap 
\begin{equation}                
\Delta^{{\cal Q}} =  E_0^{{\cal Q}}(L+1) + E_0^{{\cal Q}}(L-1) 
- 2 E_0^{{\cal Q}}(L) \, ,
\label{gapdef}
\end{equation}
and the momentum distribution
\begin{equation}
n^{{\cal Q}}_k =
\frac{1}{2} \sum_{\sigma} 
\langle \Psi_0^{{\cal Q}} |
\hat{c}_{k,\sigma}^+ \hat{c}_{k,\sigma} 
| \Psi_0^{{\cal Q}} \rangle
\, , \label{nkdef}
\end{equation}
for each configuration~${\cal Q}$, where 
$|\Psi_0^{{\cal Q}}\rangle$
is the normalized, paramagnetic 
ground-state wave function at half band-filling,
and $E_0^{{\cal Q}}(N)$ is the $N$-particle ground-state energy.
Since the self-energy is $k$-independent, 
the momentum distribution depends on~$k$ only through~$\varepsilon_k$:
$ n^{{\cal Q}}_{k} \equiv n^{\cal Q}(\varepsilon_k) $.
Thus, $Z^{\cal Q} = n^{{\cal Q}}(\varepsilon=0^-) - 
n^{{\cal Q}}(\varepsilon=0^+)$
defines the quasiparticle weight for 
configuration~${\cal Q}$
at half band-filling.

To obtain the mean value of a physical quantity,
we average over $R_L$ selected
configurations~${\cal Q}$, e.g.,
$\Delta(U;L)= (1/R_L)\sum_{\cal Q} \Delta^{\cal Q}(U;L)$.
Random sampling determines the mean value with accuracy
$\delta\Delta(U;L)= \sigma_{\Delta}(U;L)/\sqrt{R_L}$, 
where 
$\left[\sigma_{\Delta}(U;L)\right]^2 = (1/R_L) \sum_{{\cal Q}}
[\Delta^{{\cal Q}}(U;L)- \Delta(U;L)]^2$
denotes the width of the distribution for given~$U$ and~$L$.
Finally, we must
extrapolate the result to infinite system sizes,
e.g., $\Delta(U)=\lim_{L\to\infty} \Delta(U;L)$.

Here we present results for a semielliptic density of states,
$D(\varepsilon)=4/(\pi W) [1-\left(2\varepsilon/W\right)^2]^{1/2}$ 
for $|\varepsilon|< W/2$. 
In one dimension this density of states
translates into a dispersion relation which obeys the implicit relation
$k/2 = \left(2\varepsilon_k/W\right)
[1-\left(2\varepsilon_k/W\right)^2]^{1/2}
+\arcsin\left(2\varepsilon_k/W\right)$.
In the following we measure energies in units of $t=W/(2\pi)$.

We choose a particular realization  $\varepsilon_k^{{\cal Q}}$ 
for a configuration~${\cal Q}$ as a permutation
of the $L$ energy values $\left\{ \varepsilon_k \right\}$
over the momenta~$\{k\}$.
This choice avoids fluctuations in the bare
density of states.
We then select those permutations~${\cal Q}$ 
for which the joint density of states 
$D_q^{{\cal Q}}(\varepsilon_1,\varepsilon_2)$ (approximately) factorizes.
For small finite systems, we analyze the distribution of hopping
matrix elements
$|t^{\cal Q}(r)|^2 = (1/L)\sum_q \exp(iqr) 
\sum_k \varepsilon^{\cal Q}_k\varepsilon^{\cal Q}_{k+q}$.
In the thermodynamic limit we find
$|t^{\text{RDA}}(r)|^2 = \overline{t^2} =
(1/L)\sum_k \varepsilon_k^2$, independent of~$r$.
We exclude configurations~${\cal Q}$ which strongly deviate
from this relation, retaining about every second (randomly chosen)
configuration.
For $L = 4,5,6$, we calculate all unique configurations;
we select $R_L=500$ for $L = 7,8,9$, $R_L=200$ for $L = 10,11,12$,
and $R_L=50$ for $L=13, 14$.
In order to improve the statistics for small system sizes, we choose
independent random dispersions for spin-$\uparrow$~and
spin-$\downarrow$~electrons.
We emphasize that the $L\to\infty$ extrapolation does not depend
strongly on how the configurations $R_L$ are chosen;
our selection criteria simply reduce the statistical errors in order to
make the extrapolation more accurate.

\begin{figure}
\vspace*{-0.7cm}
\begin{center}
  \epsfig{file=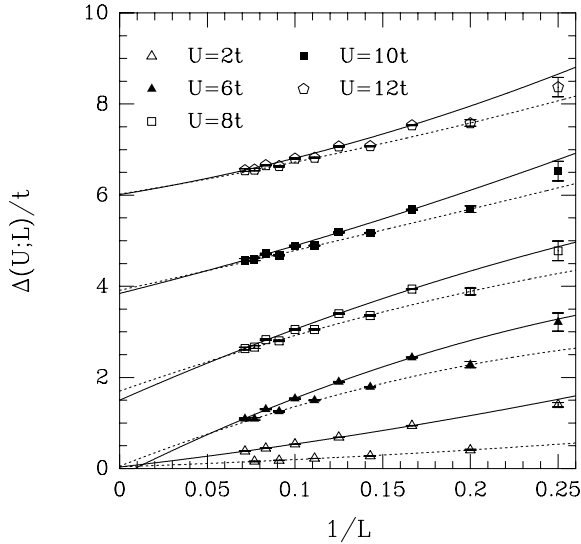,width=8cm}
\end{center}
\vspace*{-0.2cm}
\caption{One-particle gap $\Delta(U;L)$ plotted as a function of $1/L$
  for various $U$ values.
  Separate extrapolations are done for even $L$ (solid lines) and odd
  $L$ (dotted lines).
  }
\label{fig:gapL}
\end{figure}

In Fig.~\ref{fig:gapL} we plot the one-particle gap $\Delta(U;L)$, 
Eq.~(\ref{gapdef}), 
as a function of~$1/L$ for various values of~$U$.
The lines are separate ${\cal O} (1/L^2)$ extrapolations for odd and 
even lattice sites.
The bars on the symbols give the ``error'' in the mean values, 
$\delta\Delta(U;L)$. 
The statistical width introduced by the averaging procedure 
is small enough so that finite-size corrections are the dominant
effect.

The gap $\Delta(U)$,
Fig.~\ref{fig:Zgapinf}, is zero to within
the error of the finite-size extrapolation (approximately the size of
the points or less) for $U < U_{\text{c}} \approx W$.
As can be shown analytically in the large-$U$ limit~\cite{Schmit},
the upper and lower Hubbard bands have width~$W$ 
for the case of a semielliptic density of states.
In fact, above the transition we find
$\Delta(U\gtrsim 1.1W) = U-W$, with an accuracy of better than
a few percent. 
An (almost) linear behavior for $U\gtrsim 1.2 W$
was also reported in Refs.~\cite{Moeller,Logansheros}.
Since $1/U$-corrections do not contribute down to $U=1.1W$,
and finite-size corrections are largest below that value,
we conjecture that $\Delta(U)=U-W$ might hold for $U\geq U_{\text{c}}=W$.
Allowing for the possibility of two critical interaction strengths,
we give $U_{\text{c},1}=(1.0 \pm 0.1)W$ as a conservative estimate.

\begin{figure}
\vspace*{-0.7cm}
  \begin{center}
    \epsfig{file=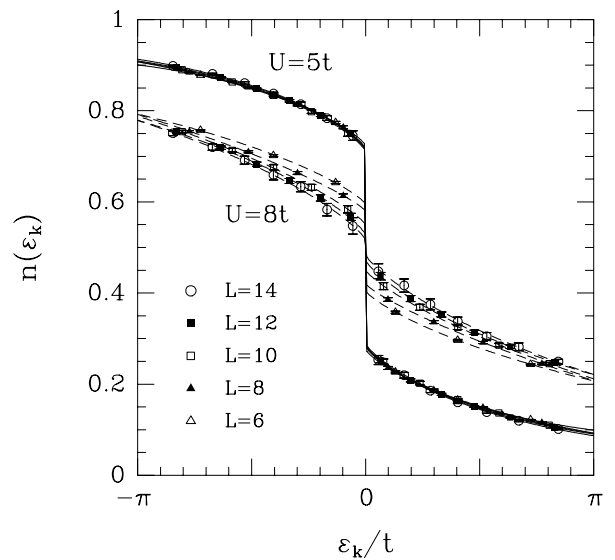,width=8cm}
  \end{center}
\vspace*{-0.2cm}
\caption{
  The energy-shell-averaged momentum distribution $n(\varepsilon_k)$ for
  $U=5t$, and $U=8t$.
  The results for $L=6,8,10,12,14$ are superimposed.
  }
\label{fig:nepsk}
\end{figure}

\begin{figure}
\vspace*{-0.8cm}
  \begin{center}
    \epsfig{file=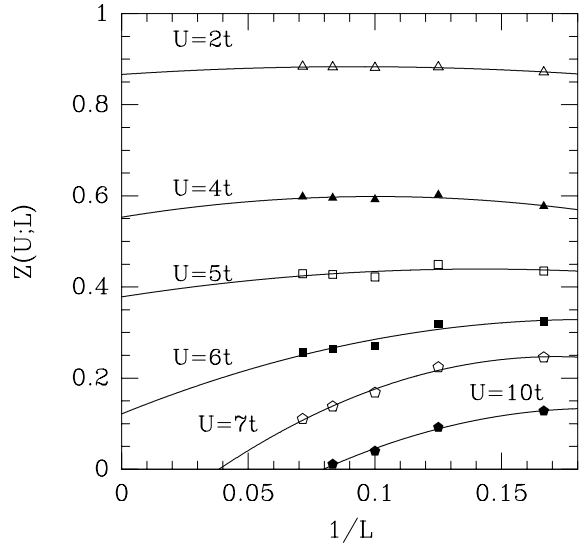,width=8cm}
  \end{center}
\vspace*{-0.3cm}
\caption{
  The quasiparticle weight $Z(U;L)$ plotted versus $1/L$ for various 
  $U$ values, fitted with a second-order polynomial in $1/L$.
  }
\label{fig:Z}
\end{figure}

In order to determine the quasiparticle weight 
$Z(U)$, we need to resolve the momentum distribution
in the vicinity of the Fermi energy.
For even system sizes the closest point $\varepsilon_k$ to the Fermi energy
$E_{\text{F}}=0$ is $\varepsilon_{(\pi/2 - \pi/L)}$. For odd system sizes
there is one level right at the Fermi energy but
the next one lies at $\varepsilon_{(\pi/2-2\pi/L)}$.
Therefore, we expect better results for $Z(U)$ from an extrapolation
of the even-size systems.
In Fig.~\ref{fig:nepsk} we display the momentum
distribution $n(\varepsilon_k)$ for two $U$~values. 
The lines are fits to the Fermi-liquid form
$n(\varepsilon<0;U;L)= [1+Z(U;L)]/2 +
\alpha(U;L) (\varepsilon/W) \ln \left(\varepsilon/W\right)
+ \beta(U;L) (\varepsilon/W) + \ldots\ ]
$
with $n(-\varepsilon)=1-n(\varepsilon)$ due to particle-hole symmetry.
In Fig.~\ref{fig:Z}, we display $Z(U;L)$, fitted by a
second order polynomial in $1/L$.
The $(1/L)\to 0$ extrapolated value, $Z(U)$, is displayed in 
Fig.~\ref{fig:Zgapinf}.

We expect finite-size effects to be under good control
if, in the interacting density of states,
we can detect the quasiparticle resonance, which, at half band-filling, 
represents $ZL/2$ electrons. To obtain a rather conservative bound we assume
that we miss the resonance if it contains less than one
electron in our largest system, $Z_{\text{min}} =2/L_{\text{max}}=1/7$.
If we further assume a reasonably regular behavior of $Z(U)$ for 
$U>0.9 W$ we find $U_{\text{c},2}=(1.0 \pm 0.1)W$ for
the vanishing of the quasiparticle weight. 
This idea is supported by the fact that fourth-order perturbation
theory, $Z^{(4)}(U) = 1 - 1.308 (U/W)^2 + 0.6369 (U/W)^4 - \ldots\ $, 
quantitatively agrees with our results up to $U \approx 0.6 W$,
above which higher-order corrections become important.

\begin{figure}
\vspace*{-0.9cm}
  \begin{center}
    \epsfig{file=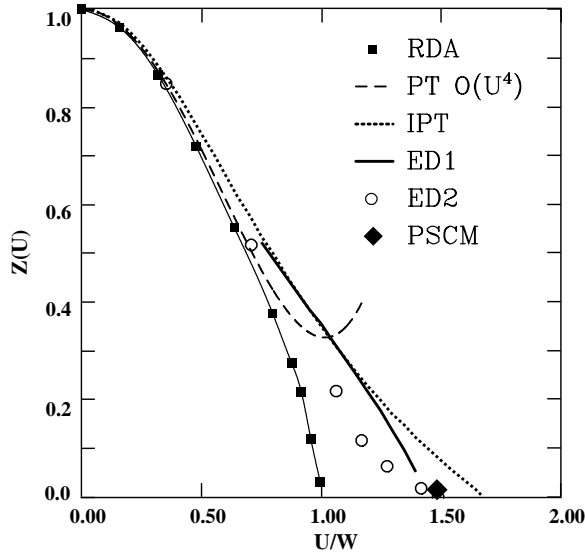,width=8cm}
  \end{center}
\vspace*{-0.3cm}
\caption{
  The quasiparticle weight $Z(U)$ within the RDA, fitted with a spline 
  function and compared to various other approaches
  described in the text.
  Also included is $U_{\text{c,2}}$ obtained from the projective
  self-consistent method (PSCM).
  }
\label{fig:summary}
\end{figure}

This comparison is shown in Fig.~\ref{fig:summary}, along with the
results from  
iterated perturbation theory (IPT), the projective self-consistent
method (PSCM) \cite{MEGAREVMOD,PSCM}, and two exact diagonalization
studies (ED1~\cite{Moeller} and ED2~\cite{CaffKrauth}) of the
self-consistency equation.
Note that, 
aside from the standard fourth-order perturbation theory
[PT ${\cal O}(U^4)$], 
all of these results significantly deviate from 
the RDA results at $U \gtrsim W/2$.

The IPT is a perturbative approach which may fail
in the vicinity of the quantum phase transition~\cite{MEGAREVMOD}.
It is seen to be quantitatively
unreliable even at moderate interaction strengths.
The PSCM also predicts a continuously
vanishing quasiparticle weight and a jump discontinuity in the gap.
It requires the separation of high and low
energy scales close to the transition. However, this assumption is in 
conflict with the mapping of the infinite-dimensional
Hubbard model onto the dynamical mean-field equations~\cite{Kehrein}.
In the ED studies of the self-consistency equations,
the energy levels of the effective medium
are discretized, which limits the energy resolution to 
$\Delta\omega\gtrsim 0.2W$ for $n_s\leq 10$ bath levels~\cite{CaffKrauth}.
This energy resolution may be insufficient
to distinguish the spectral weight of the quasiparticle resonance
from spectral weight in the tails of the Hubbard bands accurately.
The size of the quasiparticle weight and therefore the
stability of the metal could thereby be overestimated.

In this work we have applied the random dispersion approximation to 
calculate directly the gap and the momentum distribution
on finite systems. 
We have used conventional polynomial fits in $1/L$
to extrapolate to the thermodynamic limit, and have 
found results in full accordance with 
Mott's view of a {\sl continuous\/} Mott-Hubbard transition with a
single critical interaction strength.

Helpful discussions with D.\ Baeriswyl, R.\ Bulla, D.\ Logan, 
P.\ No\-zi\`eres, D.\ Vollhardt, P.G.J.\ van Dongen,
S.R.\ White and X.\ Zotos are gratefully acknowledged.

\end{multicols}

\end{document}